\documentclass{aa}
\usepackage[varg]{txfonts}
\usepackage{graphicx}

\begin{document}

\title{Distinguishing between symbiotic stars and planetary nebulae}


\author{Krystian I\l{}kiewicz \and Joanna Miko\l{}ajewska}

\titlerunning{Distinguishing between SySt and PNe}

\offprints{K. I\l{}kiewicz, \email{ilkiewicz@camk.edu.pl}}

\institute{Nicolaus Copernicus Astronomical Center, Polish Academy of Sciences, ul. Bartycka 18, 00-716 Warsaw, Poland}

\date{Received ... / Accepted ...}

\abstract {Number of known symbiotic stars (SySt) is still significantly lower than their predicted population. One of the main problems in finding complete population of SySt is the fact that their spectrum can be confused with other objects, such as planetary nebulae (PNe) or dense \ion{H}{II} regions. The problem is reinforced by a fact that in significant fraction of established SySt the emission lines used to distinguish them from other objects are not present.} {We aim at finding new diagnostic diagrams that could help separate SySt from PNe. Additionally, we examine known sample of extragalactic PNe for candidate SySt.} {We employed emission line fluxes of known SySt and PNe from the literature. } {We found that among the forbidden lines in the optical region of spectrum, only the [\ion{O}{III}] and [\ion{N}{II}] lines can be used as a tool for distinguishing between SySt and PNe, which is consistent with the fact that they have the highest critical densities. The most useful diagnostic that we propose is based on \ion{He}{I} lines which are more common and stronger in SySt than forbidden lines. All these useful diagnostic diagrams are electron density indicators that better distinguishes PNe and ionized symbiotic nebulae. Moreover, we found six new candidate SySt in the Large Magellanic Cloud and one in M81. If confirmed, the candidate in M81 would be the furthest known SySt thus far.} {}

\maketitle

\section{Introduction}
Symbiotic stars (SySt) are interacting binaries with the longest orbital periods. In these systems an evolved, cool star is transferring mass to a much hotter, and more luminous compact companion. The mass accretor is typically a white dwarf (WD), but in some cases a neutron star (NS) is observed. The mass donor is a normal red giant (RG) in S-type (stellar) SySt, or a Mira surrounded by a warm dust shell in D-type (dust) systems. In D'-type (dusty) SySt there is a F- or G-type cool giant surrounded by a dust shell. SySt are good laboratories for binary interaction and evolution because they show such phenomena as jets, accretion/excretion discs, nova outburst and interacting winds. Moreover, they are promising candidates for progenitors of type Ia supernovae (see e.g. \citealt{1999ApJ...522..487H}; \citealt{2009MNRAS.396.1086L}; \citealt{2010ApJ...719..474D}; \citealt{2011ApJ...735L..31C}).  Most recent review of properties of these systems is presented by \citet{2012BaltA..21....5M}.

Thus far $\sim$300 SySt have been discovered in the Milky Way \citep[MW][and references therein]{2000A&AS..146..407B,2013MNRAS.432.3186M,2014MNRAS.440.1410M,2014A&A...567A..49R,2015RAA....15.1332L,2016AJ....151..100B}. This number is considerably lower than the predicted number of SySt in the Galaxy, ranging from 3000 \citep{1984Ap&SS..99..101A} up to $4\times10^5$ \citep{2003ASPC..303..539M}. This, combined with the fact that the distance to most of the Galactic SySt is not known, hinders study of their properties. The situation is improving thanks to growing number of known SySt in the Local Group of Galaxies \citep[LGGS, ][]{2000A&AS..146..407B,2008MNRAS.391L..84G,2009MNRAS.395.1121K,2014MNRAS.444L..11M,2014MNRAS.444..586M,2015MNRAS.447..993G,2017MNRAS.465.1699M}, which in future may lead to observing a full sample of SySt in one of the galaxies.

Definition of a SySt that is used in literature includes the presence of a late-type giant features in the spectrum, strong emission lines of \ion{H}{I} and \ion{He}{I} as well as the presence of emission lines with a high ionization potential  \citep{1986syst.book.....K}. \citet{2017MNRAS.465.1699M} have recently highlighted that such a method of classification of SySt in LGGS can be hindered by contamination of the spectrum by diffuse interstellar gas (DIG). Moreover, in D-type systems, where the molecular bands of the RG cannot be detected, the SySt can be mistaken with a planetary nebula (PN). In fact, big fraction of the SySt have been originally classified as PNe \citep[see e.g.][and references therein]{2000A&AS..146..407B}. Thus far the most reliable criterion for SySt was the presence of \ion{O}{VI} Raman scattered 6825$\AA$ and 7082$\AA$ lines. However, this line is present only in $\sim$50\% of SySt \citep{1980MNRAS.190...75A}. 

\citet{1995PASP..107..462G} proposed a diagnostic diagram using [\ion{O}{III}] and Balmer emission lines, that successfully have been employed to classify new SySt.  The [\ion{O}{III}] lines are not always detected in SySt, so the \citet{1995PASP..107..462G} criterion is not always applicable. Moreover, some SySt have been observed in the region dominated by young PNe \citep{2010IAUS..262..307B}. In some studies criteria based on infrared colors have been suggested \citep{2001A&A...377L..18S,2014A&A...567A..49R}, however these criteria are only applicable in the Galaxy. Because of the above restrictions,  in this paper, we explore possibility of using other emission lines for classifying new SySt. Similar comparisons of the emission line ratios in SySt and PNe had been carried out in the past \citep[e.g.][]{1986A&A...166..143G}, however to our knowledge this is the first study focussed on new criteria and using such a big sample of data. Moreover, in this study we examine known sample of extragalactic PNe in order to detect misclassified SySt among them. 

\section{Observations}\label{obs_sec}

\begin{figure*}
  \resizebox{0.5\hsize}{!}{\includegraphics{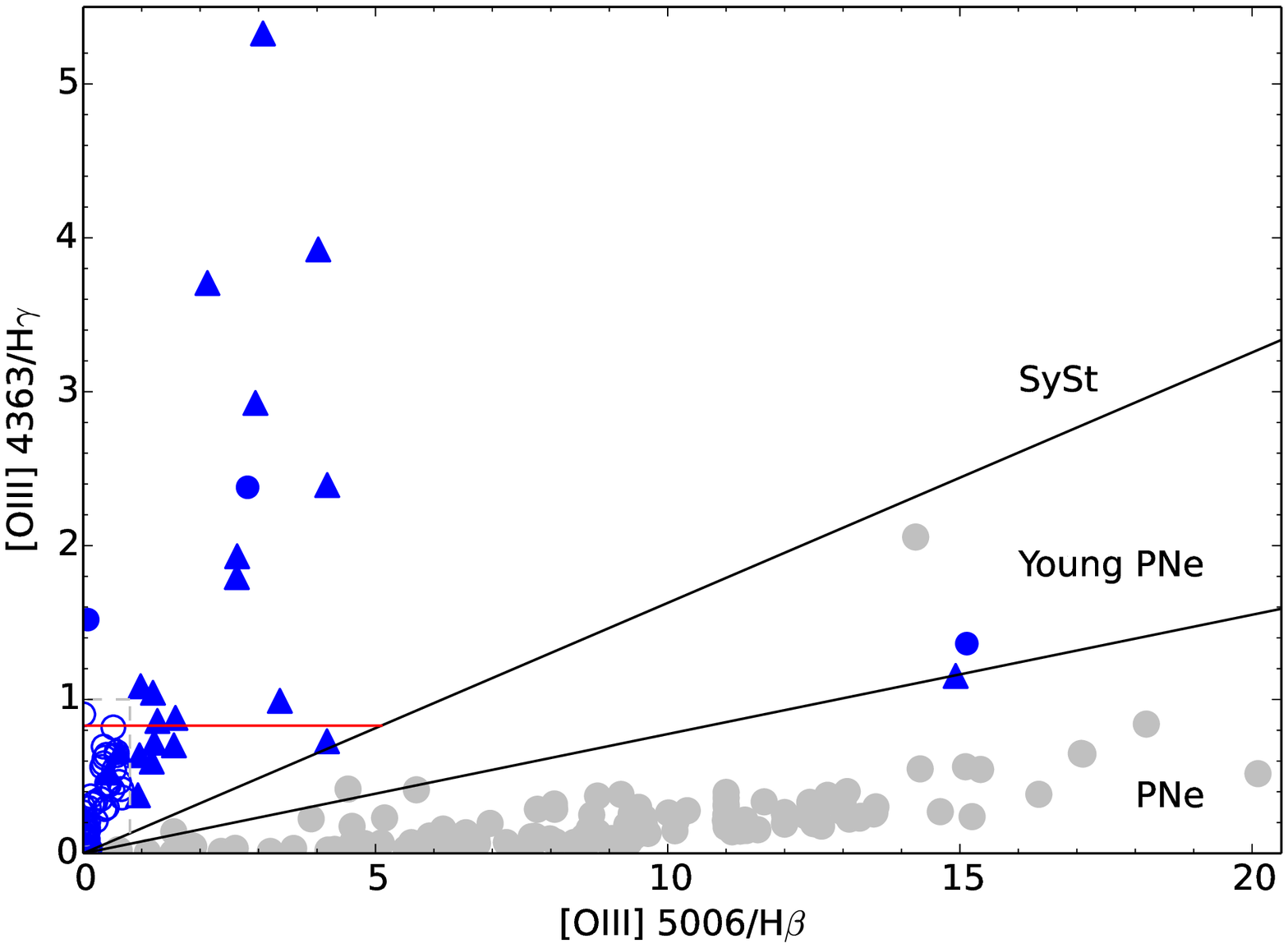}}
  \resizebox{0.5\hsize}{!}{\includegraphics{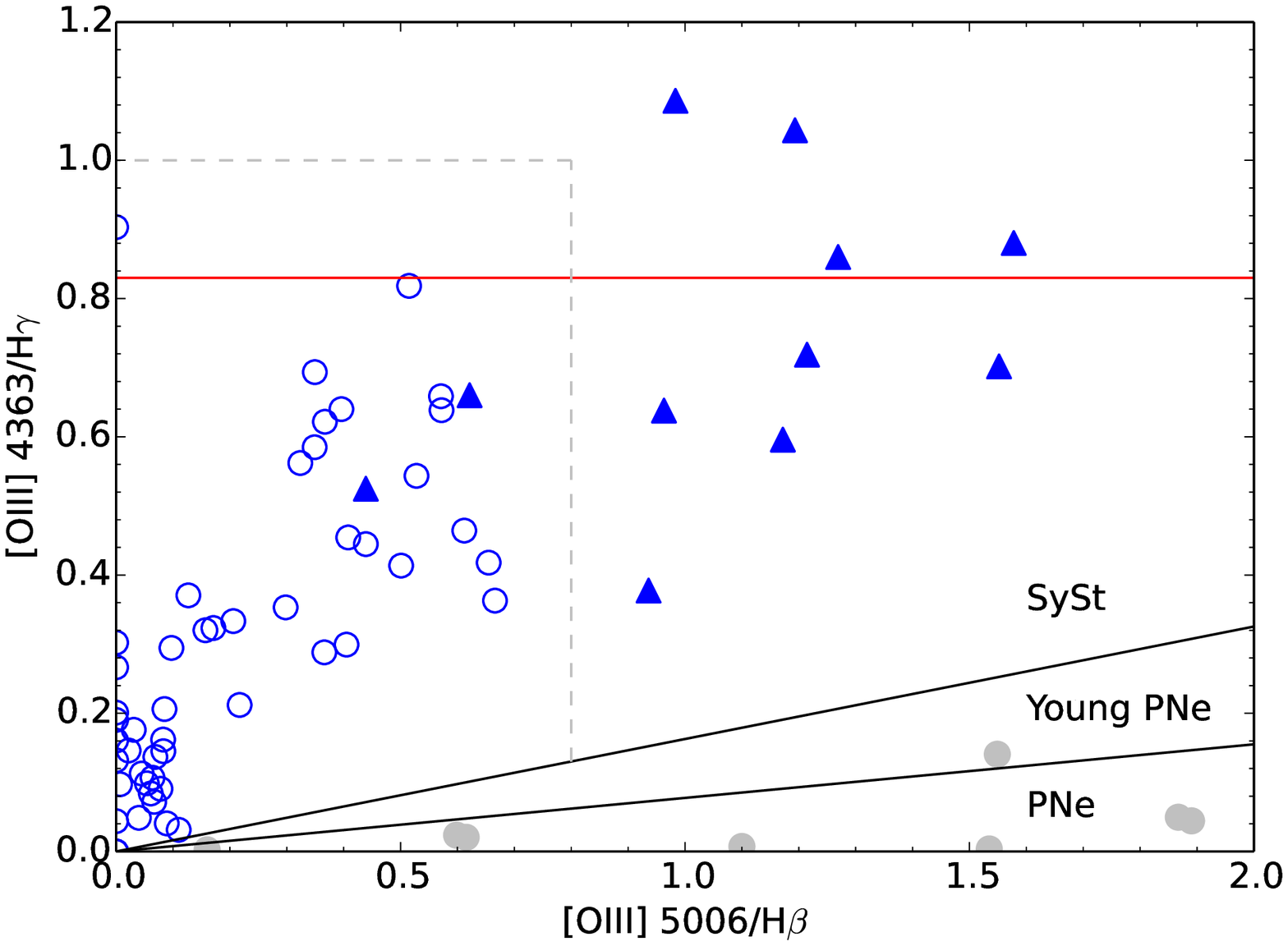}}
  \caption{The [\ion{O}{III}] diagnostic diagram of \citet{1995PASP..107..462G} for SySt (blue symbols) and PNe (gray symbols) in the MW. The S-type SySt are marked with open circles, the D'-type SySt with full circles, and D-type SySt with triangles. The red line marks  \citet{2008MNRAS.384.1045K} criterion for distinguishing between S- and D-type SySt. The gray lines indicate the regions of S- and D-type SySt proposed in this work.  Left and right panel are the same diagram, but with different zoom. }
  \label{oiii_MW}
\end{figure*}

\begin{figure*}
  \resizebox{\hsize}{!}{\includegraphics{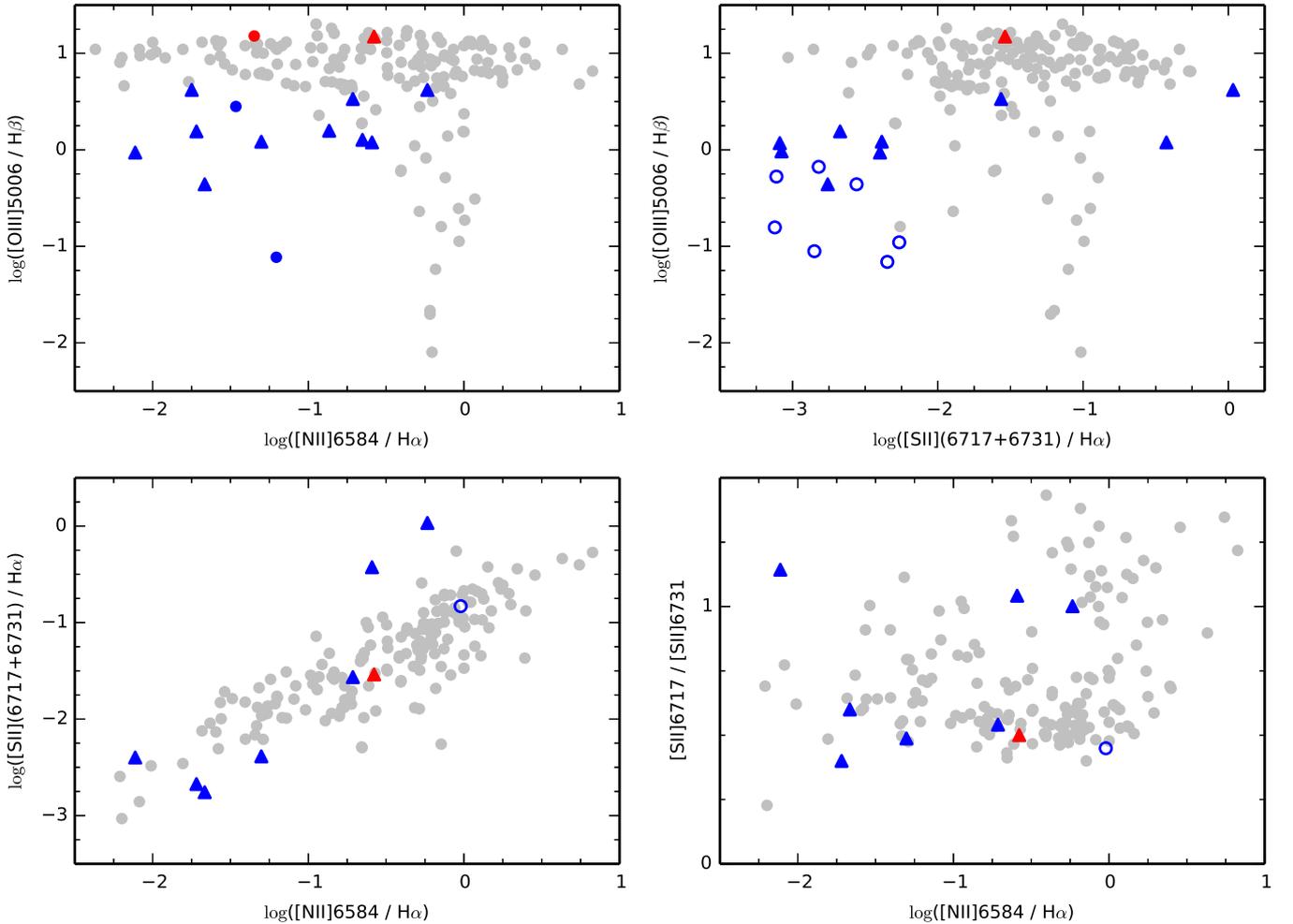}}
  \caption{\citet{2008MNRAS.384.1045K} diagrams used for distinguishing between PNe and \ion{H}{II} regions. The SySt are marked with blue symbols and PNe with gray symbols. The SySt that  fall into the PNe region in the [\ion{O}{III}] diagnostic diagram are marked with red symbols. The S-type SySt are marked with open circles, the D'-type SySt with full circles, and D-type SySt with triangles. }
  \label{diagnostic_kniazev}
\end{figure*}

\begin{figure*}
  \resizebox{\hsize}{!}{\includegraphics{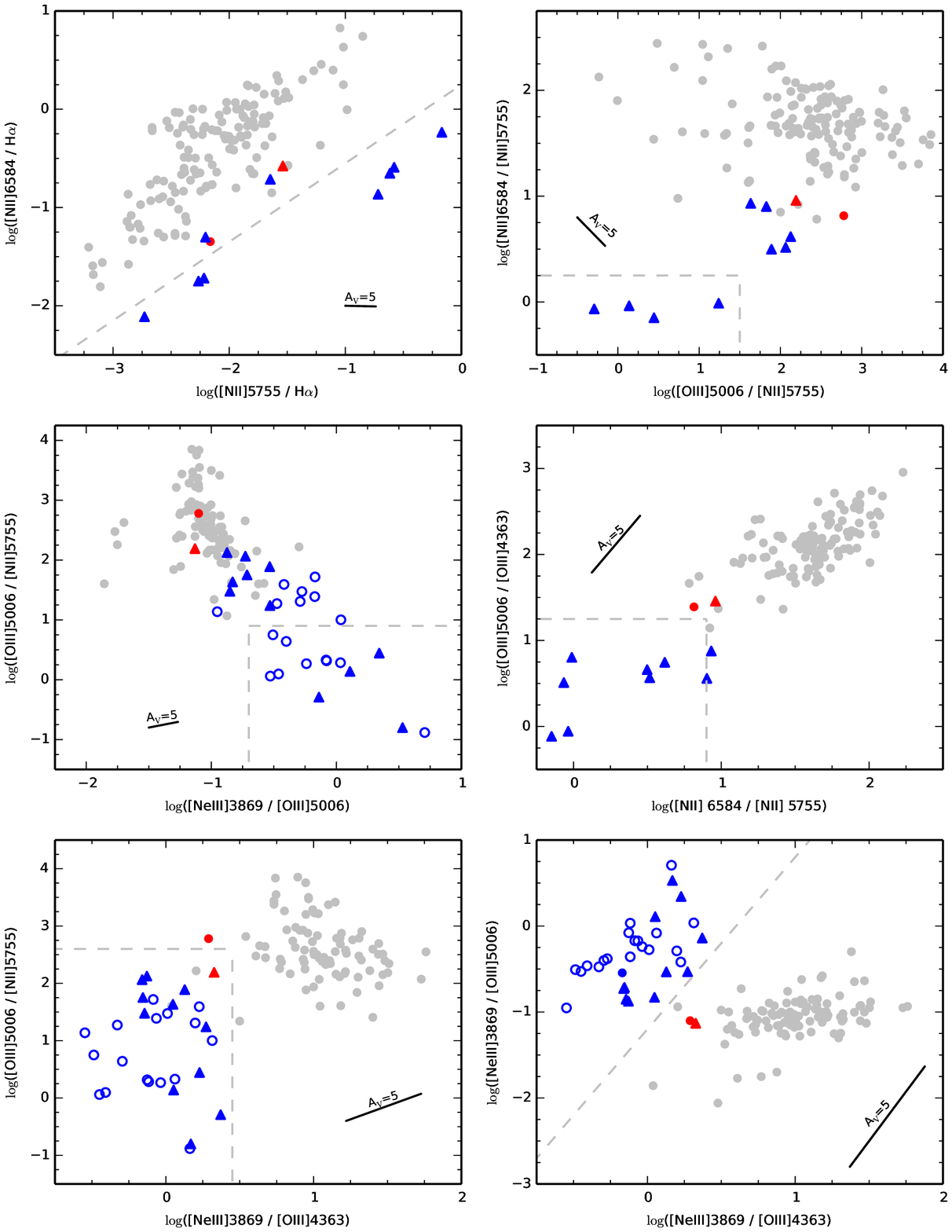}}
  \caption{Various emission line ratios relationships for Galactic SySt and PNe employing forbidden line fluxes. Symbols are the same as in Fig.~\ref{diagnostic_kniazev}. The reddening was calculated using \citet{1989ApJ...345..245C} extinction model with R$_V$=3.1 assumed. The gray lines indicate new criteria for SySt proposed in this work.}
  \label{diagnostic_other}
\end{figure*}

In order to examine possibility of new diagnostics for SySt we used emission line fluxes of PNe and SySt from the MW, where the objects are well studied and  contamination of the spectra by DIG is usually not a concern.  In the case of PNe, the emission line fluxes  were taken from \citet{2004A&A...427..231G}, \citet{2009A&A...500.1089G} and \citet{2010ApJ...724..748H}. Among the objects in these catalogs we reclassified known or proposed SySt: \object{PN ShWi 5} \citep{2009A&A...496..813M},	\object{PN H 2-43} \citep{2000A&AS..146..407B}, \object{PN G356.9-05.8} \citep{2003A&A...404..545Z}, and \object{PN G005.2+04.2} \citep{2009A&A...496..813M}.  Fluxes of emission lines in SySt were taken from \citet{1997A&A...327..191M}, \citet{1998PASP..110..458G}, \citet{1998A&A...333..658P} and \citet{2005A&A...435.1087L}. The D-, S- and D'-type classification of individual stars was adopted from \citet{2000A&AS..146..407B} and \citet{2007MNRAS.376.1120P}.  In our study we omitted stars without this classification available. Fluxes of emission lines of both PNe and SySt were reddening corrected using reddening estimates given in the employed catalogs. The \citet{1995PASP..107..462G} diagnostic diagram for PNe and SySt is presented in Fig.~\ref{oiii_MW}. Only two SySt would be misclassified as PNe based on this diagram, He~2-172 and H~1-36. Similar diagrams, using other emission lines are presented in Fig.~\ref{diagnostic_kniazev}, Fig.~\ref{diagnostic_other} and Fig.~\ref{He_diagram}.

We investigated known PNe for misclassified SySt in five members of LGGS. In the case of the Small and Large Magellanic Clouds (SMC and LMC, respectively) we employed emission line fluxes reported by \citet{1988MNRAS.234..583M} and \citet{2006A&A...456..451L}. In the case of M81 the \citet{2010A&A...521A...3S} measurements were used. For NGC300 the fluxes were taken from \citet{2013A&A...552A..12S}. he M33 data is from \citet{2009ApJ...696..729M}.  The position of extragalactic PNe on the [\ion{O}{III}] diagnostic diagram is presented in Fig.~\ref{PNe_cands}.

\section{Results}\label{res_sec}

\subsection{D- and S-type classification}

Most of the SySt are clearly separated from PNe in the [\ion{O}{III}] diagram (Fig.~\ref{oiii_MW}). The exception are two SySt - He~2-172 and H~1-36. Using the \citet{1995PASP..107..462G} criterion for separation of D- and S-type SySt eight D-type SySt would be misclassified as well as one S-type SySt. Hence, we propose a new criterion for classifying SySt types, namely when [\ion{O}{III}]~4363/H$\gamma<1$ and [\ion{O}{III}]~5006/H$\beta<0.8$ the SySt is a S-type system. With this new criterion only two D-type SySt would be misclassified and all of S-type SySt would be classified correctly.

\subsection{Distinguishing between SySt and PNe}

On the \citet{2008MNRAS.384.1045K} diagnostic diagrams, used for separating PNe and \ion{H}{II} regions, the SySt occupy the same region as PNe (Fig.~\ref{diagnostic_kniazev}). This is not surprising given the relatively low critical densities N$_{cr}$ of the [\ion{S}{II}]~6717, [\ion{S}{II}]~6731 and [\ion{N}{II}]~6584 lines. Namely, $\log(\mathrm{N}_{cr}/\mathrm{cm}^{-3})=3.2$ for [\ion{S}{II}]~6731 line,  $\log(\mathrm{N}_{cr})=3.6$ for [\ion{S}{II}]~6731 and $\log(\mathrm{N}_{cr})=4.9$  for [\ion{N}{II}]~6584 line, which is significantly lower than  $\log(\mathrm{N}_{cr})=5.8$ for [\ion{O}{III}]~5006 and $\log(\mathrm{N}_{cr})=7.5$ for [\ion{O}{III}]~4363 lines \citep{1988AJ.....95...45A}. The only  promising diagram is the diagram including [\ion{O}{III}] and [\ion{S}{II}] lines in which there is a region occupied only by Galactic SySt. However, this region seems to be occupied by extragalactic PNe and \ion{H}{II} regions \citep[see fig.~3 of][]{2008MNRAS.384.1045K}. The difference in line ratios is probably caused by lower metallicity of objects in \citet{2008MNRAS.384.1045K} sample, which caused lower [\ion{O}{III}] and [\ion{S}{II}] line fluxes compared to \ion{H}{I} lines.

We checked various other line ratios and most of the studied diagrams were not helpful in distinguishing between PNe and SySt. The most promising diagrams are presented on Fig.~\ref{diagnostic_other}.  Most of the diagrams included both [\ion{O}{III}] lines, which could be helpful when H$\beta$ and H$\gamma$ cannot be measured for the \citet{1995PASP..107..462G} diagram (Fig.~\ref{oiii_MW}). Most notably, diagrams with [\ion{N}{II}]~5755 line are the only diagrams without [\ion{O}{III}] lines in which there are regions where only SySt are present (Fig.~\ref{diagnostic_other}). Separation of SySt from PNe could be expected given [\ion{N}{II}]~5755 line critical density similar as for [\ion{O}{III}] lines, namely $\log(\mathrm{N}_{cr})=7.5$ for [\ion{N}{II}]~5755 line \citep{1988AJ.....95...45A}. 

\begin{figure}
  \resizebox{\hsize}{!}{\includegraphics{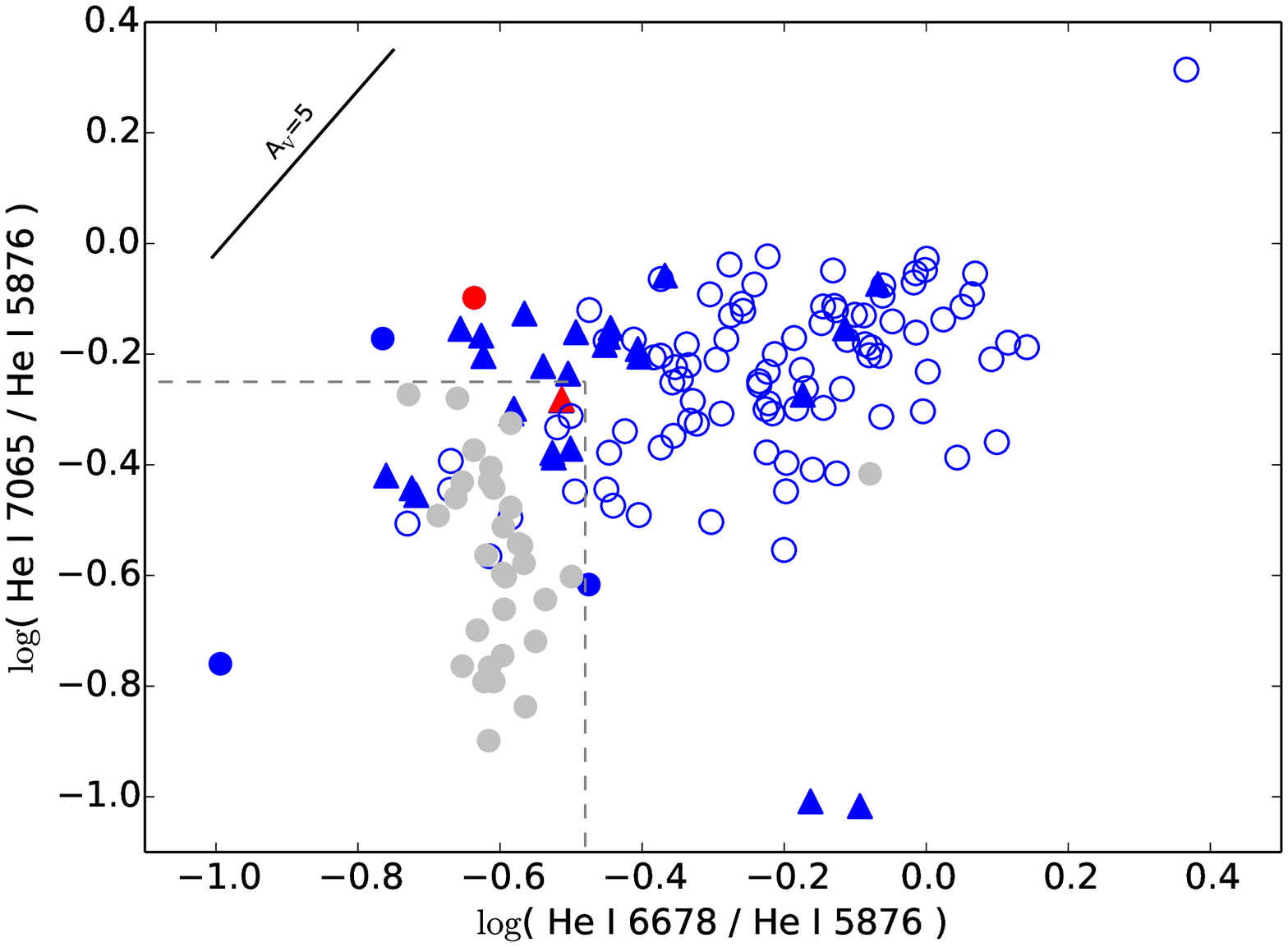}}
  \caption{Same as Fig.~\ref{diagnostic_other}, but for \ion{He}{I} line ratios.}
  \label{He_diagram}
\end{figure}

The diagnostic diagrams discussed above are all based on forbidden lines, which can help to distinguish between SySt and PNe due to the fact that they are good electron density indicators. Another kind of diagnostic diagram, not based on forbidden lines, is the one employing \ion{He}{I} lines (Fig.~\ref{He_diagram}). The \ion{He}{I} emission line ratios in SySt deviate from Case~B significantly due to metastability of the 2$^3$S level, and they can serve as a diagnostic of density. Thereby, the \ion{He}{I} emission line ratios were used in the past to distinguish between D- and S-type SySt \citep{1994MNRAS.268..213P}. In our sample of 178 PNe, only for 31 PNe the \ion{He}{I} fluxes were available. The PNe occupy only a small region in the \ion{He}{I} diagram (Fig.~\ref{He_diagram}), while SySt can be found in much bigger region. The only outlier from the PNe region (Fig.~\ref{He_diagram}) out of the 31 PNe with measured \ion{He}{I} lines is \object{PN K 3-90}.  Out of 123 SySt only 45 can be found in the region occupied by PNe. This shows that the \ion{He}{I} diagram can be used as a diagnostic diagram. Moreover, in some SySt the \ion{He}{I} emission line ratios are correlated with the orbital phase \citep[e.g.][]{2015MNRAS.451.3909I}, so some SySt can possibly be found in the region occupied by PNe only at certain orbital phase. Furthermore, in some SySt \ion{He}{I} lines can be observed only during active phases \citep[e.g.][]{2016MNRAS.462.2695I}. 

The \ion{He}{I} diagnostic diagram is the most useful new diagnostic diagram. While the separation between PNe and SySt is not as good as in the case of \citet{1995PASP..107..462G} diagram, one of the SySt that would be misclassified as a PNe based on [\ion{O}{III}] diagnostic diagram (Fig~\ref{oiii_MW}), He~2-172, would be correctly classified as SySt based on the \ion{He}{I} diagram. This is the only diagnostic diagram that would result in correct classification of this system (Fig.~\ref{diagnostic_other}). The other diagrams could be helpful when the reliable [\ion{O}{III}] lines fluxes are not available (e.g. due to saturation), however they are of limited use, since the [\ion{O}{III}] lines are more common than the [\ion{N}{II}] lines. All of the proposed criteria for SySt are presented in Table~\ref{table:criteria}.

\begin{table*}
\caption{Proposed criteria for SySt based on emission line ratios.}         
\label{table:criteria}     
\centering                       
\begin{tabular}{c}       
\hline\hline   
              
$\log($[\ion{N}{II}]~6584/H$\alpha)<0.8\times\log$([\ion{N}{II}]~5755/H$\alpha$)+0.25     \\  

$\log(\mathrm{[\ion{N}{II}]~6584} / \mathrm{[\ion{N}{II}]~5755})<0.25$ \textbf{and} $\log(\mathrm{[\ion{O}{III}]~5006} / \mathrm{[\ion{N}{II}]~5755})<1.5$\\

$\log(\mathrm{[\ion{O}{III}]~5006} / \mathrm{[\ion{N}{II}]~5755})<0.9$ \textbf{and} $\log(\mathrm{[\ion{Ne}{III}]~3869} / \mathrm{[\ion{O}{III}]~5006})>-0.7$\\

$\log(\mathrm{[\ion{O}{III}]~5006} / \mathrm{[\ion{O}{III}]~4363})<1.25$ \textbf{and} $\log(\mathrm{[\ion{N}{II}]~6584} / \mathrm{[\ion{N}{II}]~5755})<0.9$ \\

$\log(\mathrm{[\ion{O}{III}]~5006} / \mathrm{[\ion{N}{II}]~5755})<2.6$  \textbf{and} $\log(\mathrm{[\ion{Ne}{III}]~3869} / \mathrm{[\ion{O}{III}]~4363})<0.45$\\

$\log($[\ion{Ne}{III}]~3869/[\ion{O}{III}]~5006$)>2.0\times\log$([\ion{Ne}{III}]~3869/[\ion{O}{III}]~4363$)-1.2$     \\  

$\log(\mathrm{\ion{He}{I}~7065} / \mathrm{\ion{He}{I}~5876})<-0.25$  \textbf{and} $\log(\mathrm{\ion{He}{I}~7065} / \mathrm{\ion{He}{I}~5876})<-0.48$\\

\hline                    
\end{tabular}
\end{table*}

\subsection{SySt candidates among extragalactic PNe}

\begin{figure*}
  \resizebox{\hsize}{!}{\includegraphics{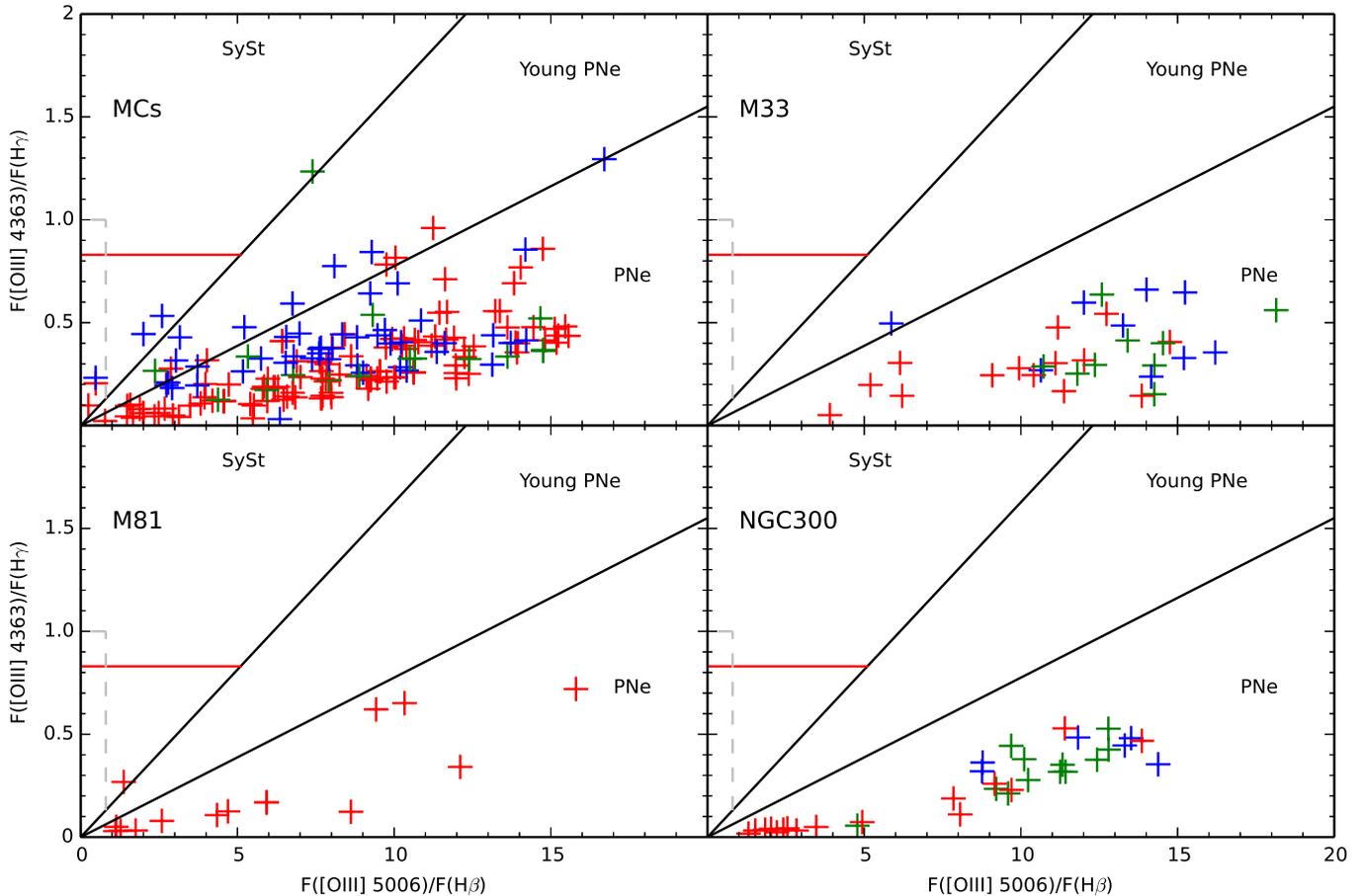}}
  \caption{The [\ion{O}{III}] diagnostic diagram for candidate SySt among extragalactic PNe with \ion{He}{II}~4686/H$\beta$ ratios $>0.3$ (blue points),  $<0.3$ (green points), and  \ion{He}{II}~4686 emission not detected (red points). The red line marks  \citet{2008MNRAS.384.1045K} criterion for distinguishing between S- and D-type SySt. The grey lines indicate the regions of S- and D-type SySt proposed in this work. }
  \label{PNe_cands}
\end{figure*}

\begin{figure}[!b]
  \resizebox{\hsize}{!}{\includegraphics{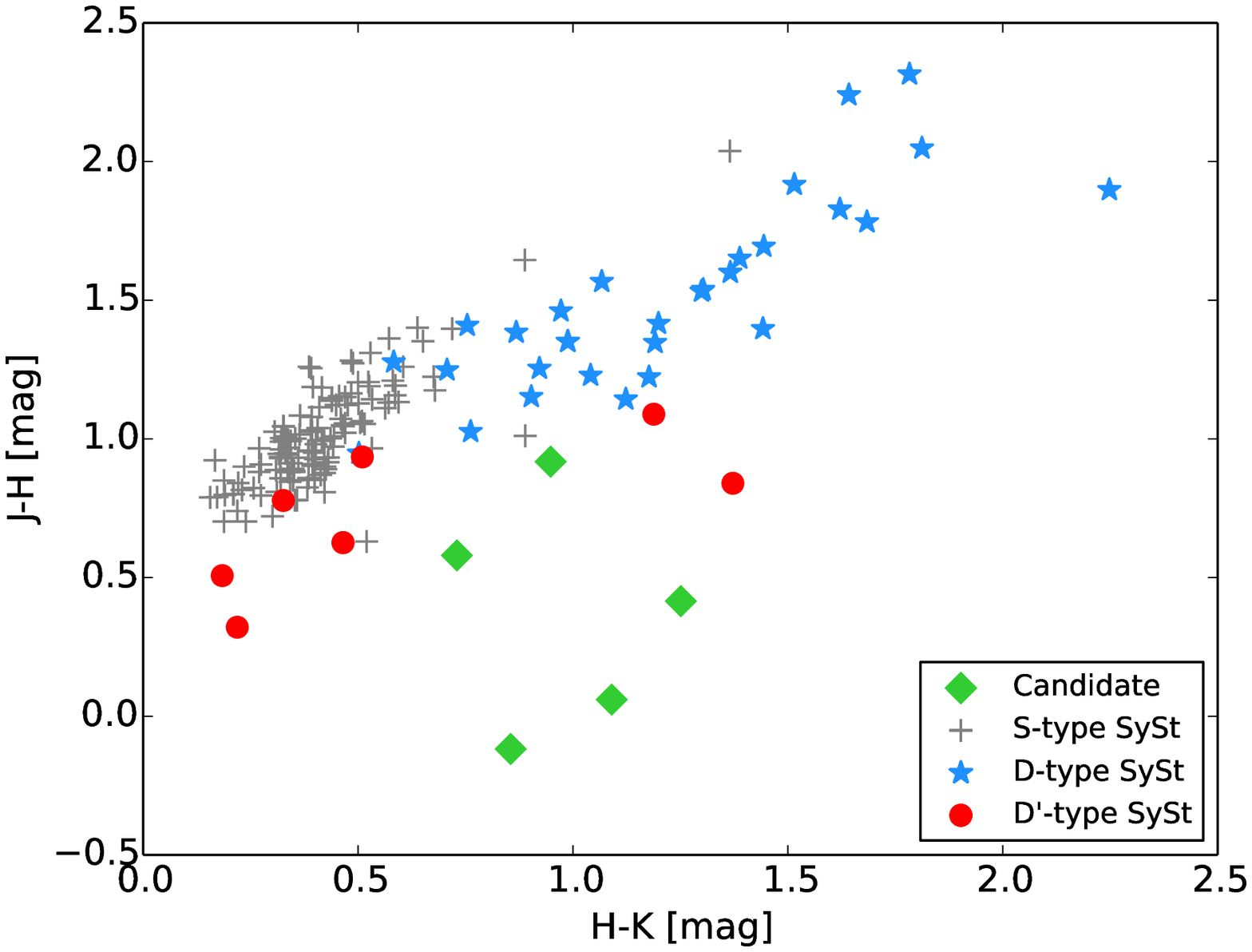}}
  \caption{Infrared colors of the LMC SySt candidates. The SySt candidates colors are from Table~\ref{table:1}. The SySt colors are from the 2MASS~6x catalog \citet{2012yCat.2281....0C}.  }
  \label{2mass_cands}
\end{figure}

Using the [\ion{O}{III}] diagnostic diagram we found six SySt candidates among the LMC PNe (Fig.~\ref{PNe_cands}). This is a high number given that only eight SySt are known in the LMC \citep{2000A&AS..146..407B}. Out of these six candidates in two of them no \ion{He}{II}~4686 line has been detected, in one the \ion{He}{II}~4686 emission is relatively weak, and in three the \ion{He}{II}~4686 emission is strong (Table~\ref{table:1}). The infrared colors (Fig.~\ref{2mass_cands}) place only two of these stars near the region occupied by SySt (\object{SMP LMC 93}, \object{SMP LMC 104}). We conclude that these two are the best candidates for SySt. The other candidates, for which the infrared colors are available, may be unusual PNe, or D'-type SySt with infrared colors more similar to those of PNe. 

\begin{figure}[!b]
  \resizebox{\hsize}{!}{\includegraphics{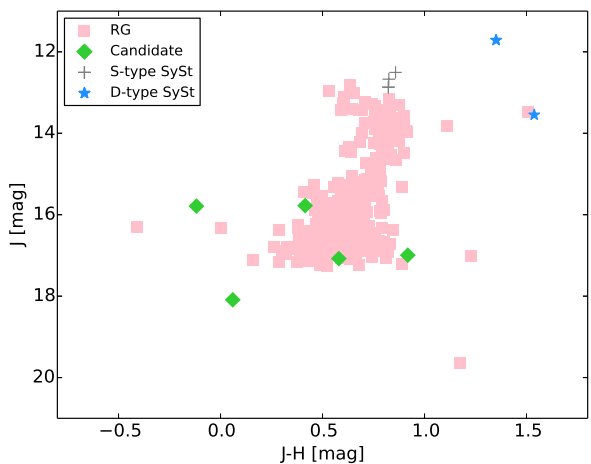}}
  \caption{Position of the LMC SySt candidates and spectroscopically confirmed RGs in the LMC on a near-infrared color-magnitude diagram. The RGs were from \citep{2011MNRAS.413..837M} and \citep{2011AJ....142...61C}. The SySt candidates colors are from Table~\ref{table:1}. The SySt and RGs colors are from the 2MASS~6x catalog \citet{2012yCat.2281....0C}.  }
  \label{2mass_cands_2}
\end{figure}

The effective temperatures of giants in SySt candidates were calculated using the formula, $T_{eff}=7070/[(J-K)+0.88]$, from \citet{1983MNRAS.202...59B}. The bolometric magnitudes were calculated using the $K$ magnitudes and bolometric corrections, $BC_K=-6.75\log (T_{eff}/9500)$, from \citet{2010MNRAS.403.1592B}. Using the LMC distance modulus of 18.493~mag \citep{2013Natur.495...76P} we were able to calculate the absolute bolometric magnitudes $M_{bol}$. The results are presented in Table~\ref{table:2}. The calculated effective temperatures are typical for SySt. On the other hand, the absolute bolometric magnitudes for most of the candidates are lower than typical SySt magnitudes \citep[see fig. 3 of][]{2007BaltA..16....1M}. However, their position on the near-infrared color-magnitude diagram is consistent with a presence of a low-luminosity RG (Fig.~\ref{2mass_cands_2}). The low luminosity of RGs in these systems could explain why they were not classified as SySt in the past. Moreover, we point out that we did not include reddening correction, the bolometric correction was sensitive to possible errors in deriving the effective temperature, and there could be error in identifying infrared counterparts of candidates.

The only candidate SySt found in other galaxies is HII403 in M81. If confirmed, this would make this object the first SySt discovered in this galaxy and the furthest known SySt \citep{2008MNRAS.391L..84G}. However, more observations are needed since it was originally classified as a \ion{H}{II} region \citep{2010A&A...521A...3S}.

We stress that deep spectra are needed to confirm nature of all of the candidates for new SySt. However, objects in the LMC with known JHK magnitudes are most likely SySt or PNe as opposed to other SySt mimics that could reproduce similar spectra and NIR colors. In particular, the most numerous SySt mimics, such as young stellar objects and T~Tauri stars, B[e] and classical Be stars as well as Wolf-Rayet stars,   \citep[see e.g.][and references therein]{2008A&A...480..409C,2010A&A...509A..41C,2014A&A...567A..49R,2014MNRAS.440.1410M} would not have similar position in the color-magnitude diagram as the SySt candidates (Fig.~\ref{2mass_cands_2}).   

\begin{table*}
\caption{Candidate SySt among the extragalactic PNe together with emission line ratios and D- and S-type classification.}         
\label{table:1}     
\centering                       
\begin{tabular}{c c c c c c c }       
\hline\hline                 
Galaxy & Name & \ion{He}{II}~4686/H$\beta$ & [\ion{O}{III}]~4363/H$\gamma$ & [\ion{O}{III}]~5006/H$\beta$ &  Ref. & Type\\    
\hline                           
LMC & \object{SMP LMC 16}  & 0.30 & 1.23  & 7.40 & 1 & D \\  
LMC & \object{SMP LMC 31}  &      & 0.21  & 0.48 & 2 & S \\   
LMC & \object{SMP LMC 63}  &      & 0.10  & 0.24 & 1 & S \\
LMC & \object{SMP LMC 93}  & 0.62 & 0.23  & 0.48 & 1 & S \\
LMC & \object{SMP LMC 104} & 0.62 & 0.44  & 2.00 & 1 & D \\
LMC & \object{MGPN LMC 31} & 0.64 & 0.53  & 2.59 & 1 & D \\
M81 & HII403               &      & 0.28  & 1.37 & 3 & D \\
\hline                    
\end{tabular}
\tablebib{(1)~\citet{2006A&A...456..451L}; (2)~\citet{1988MNRAS.234..583M}; (3)~\citet{2010A&A...521A...3S}
}
\end{table*}

\begin{table*}
\caption{Infrared magnitudes of candidate SySt with inferred properties of cool giants.}         
\label{table:2}     
\centering                       
\begin{tabular}{c c c c c c c }       
\hline\hline                 
Name & $J$ [mag] & $H$ [mag] & $K$ [mag] & Ref. & $T_{eff}$ [K] & $M_{bol}$ [mag] \\    
\hline                           
\object{SMP LMC 16} & 18.09 & 18.03 & 16.94 & 1 & 3500 & 1.39 \\  
\object{SMP LMC 31} & 15.777 & 15.362 & 14.111 & 2 & 2800 & -0.78 \\   
\object{SMP LMC 63} & 15.789 & 15.907 & 15.052 & 1 & 4400 & -1.17 \\
\object{SMP LMC 93} & 17.077 & 16.497 & 15.767 & 1 & 3200 & 0.44 \\
\object{SMP LMC 104} & 16.994 & 16.076 & 15.128 & 2 & 2600 & 0.46 \\
\hline                    
\end{tabular}
\tablebib{ (1)~\citet{2012yCat.2281....0C}; (2)~\citet{2006AJ....131.1163S}
}
\end{table*}

\section{Conclusions}\label{conlc_sec}

In this work we attempted to find new diagnostic diagrams for distinguishing SySt and PNe. We also searched for candidate SySt among the known sample of extragalactic PNe. The main results of the work include:
\begin{itemize}
\item The only diagnostic diagrams that employ forbidden lines and can distinguish between SySt and PNe are the diagrams with [\ion{O}{III}] or [\ion{N}{II}] emission lines. This is due to the relatively high critical densities of these lines. Diagnostic diagram with [\ion{O}{III}] lines was already known in the literature \citep{1995PASP..107..462G}.
\item We propose diagnostic diagram with \ion{He}{I} emission lines as a new tool for distinguishing between SySt and PNe. The advantage of this diagram is that it can correctly identify SySt that would be misclassified using the standard [\ion{O}{III}] diagnostic diagram. Moreover, the \ion{He}{I} diagram was already used in the past to distinguish between S- and D-type SySt \citep{1994MNRAS.268..213P}.
\item All the diagnostic diagrams that are useful for distinguishing between SySt an PNe are electron density indicators, which is the physical property that is most useful for distinguishing the ionized nebulae of the two classes of objects.
\item We found six candidate SySt in the LMC, while only eight SySt were known in this galaxy thus far. Additionally, we found one candidate in M81, which if confirmed, would be the first SySt discovered in this galaxy, and the furthest known SySt \citep{2008MNRAS.391L..84G}. 
\end{itemize}

In future studies it would be useful to search for new diagnostic diagrams in other spectral ranges, where forbidden lines with different critical densities can be found.

\begin{acknowledgements} 
This study has been financed by the Polish Ministry of Science and Higher Education Diamond Grant Programme via grant 0136/DIA/2014/43. This study has been supported in part by the Polish National Science Centre grant DEC-2013/10/M/ST9/00086. This research has made use of the VizieR catalogue access tool, CDS, Strasbourg, France. The original description of the VizieR service was published in A\&AS 143, 23.

\end{acknowledgements}

\bibpunct{(}{)}{;}{a}{}{,} 
\bibliographystyle{aa} 
\bibliography{references} 
\end{document}